 \documentclass[final,5p,times,twocolumn]{elsarticle}
\biboptions{square,numbers,sort&compress,comma}

\usepackage{amssymb}
\usepackage{amsmath}
\usepackage{hyperref}
\usepackage{xcolor}
\usepackage{float}
\usepackage[normalem]{ulem}
\usepackage{booktabs}

\journal{Physics Letters B}

\begin{document}

\begin{frontmatter}



\title{Probing the structure of the two proton Borromean nucleus $^{17}\rm{Ne}$ through inclusive breakup data}


\author[a]{B. Pinheiro}
\author[b]{M. Rodr\'{i}guez-Gallardo}
\author[a]{J. Lubian}

 \affiliation[a]{organization={Instituto de F{\'\i}sica, Universidade Federal Fluminense},
             city={Niter\'{o}i},
             postcode={24210-340},
             state={RJ},
             country={Brazil}}

 \affiliation[b]{organization={Departamento de F\'{i}sica At\'{o}mica, Molecular y Nuclear, Universidad de Sevilla},
             city={Seville},
             postcode={E-41012},
             country={Spain}}

\begin{abstract}
The proton-rich nucleus $^{17}$Ne exhibits a Borromean structure, $^{15}$O~+~p~+~p, since both $^{16}$F and diproton systems are unbound. The low-energy spectrum of this nucleus consists of a series of resonances that remain poorly identified. In this work, a three-body model is used to describe the structure of $^{17}$Ne, with special attention to the emergence of resonant states. The four-body continuum-discretized coupled-channel (CDCC) formalism is used to study the reaction of $^{17}$Ne on a $^{208}$Pb target near the Coulomb barrier. Comparing the calculated elastic and breakup cross sections with the recently available experimental data~\cite{OVEJAS2023138007} provides an excellent framework for probing the resonant structure of this nucleus.
\end{abstract}




\end{frontmatter}




\section{Introduction}
\label{introduction}


The exploration of nuclei far from the valley of stability has been one of the main challenges in nuclear physics. As the limits of nuclear binding are approached, exotic structures emerge, such as nuclear halos. In these weakly bound systems, the wave function of one or more valence nucleons extends far beyond the core radius, forming a diffuse, extended matter distribution. This spatial extension has profound, directly observable consequences for reaction dynamics. It effectively reduces the Coulomb barrier and enhances fusion cross sections at energies below and around the barrier~\citep{canto2015recent,canto2009dynamic,rangel2026nucleus}. A particularly interesting case is Borromean nuclei, three-body systems in which removing any one constituent leaves the entire system unbound, even though the three-body system is bound.\\
Candidates for proton-rich halo nuclei are significantly less common than the neutron-rich ones. This arises from the required proton-neutron imbalance and the presence of the Coulomb barrier. For heavier systems, the Coulomb repulsion confines the proton wave function, suppressing the tunneling that gives rise to a halo. However, for light proton-rich nuclei, this Coulomb suppression is weaker, and continuum effects can be both significant and experimentally accessible. \\
Among light proton-rich nuclei, two prominent candidates for a proton halo structure have attracted strong interest: $^{8}$B and $^{17}$Ne. The $^{17}$Ne nucleus has attracted particular interest because of its Borromean structure, characterized by an inert $^{15}$O core and two valence protons with a two-proton separation energy of $\approx 0.94$ MeV. This inherently three-body character is further accentuated by the observation of two-proton decay from its low-lying states to $^{15}$O~\citep{chromik2002two,Chromik2,PhysRevC.96.025807}. This decay channel directly connects to the astrophysically important CNO cycle during explosive hydrogen-burning scenarios, such as X-ray bursts. The $^{15}$O(2p,$\gamma$)$^{17}$Ne capture reaction is, therefore, of significant astrophysical interest and requires a robust structural and reaction model~\citep{Casal2016,tanaka2010density,gorres1995bridging}. The spectroscopy of $^{17}$Ne is challenging, with the controversy concerning its low-lying spectrum, particularly the $1/2^{+}$ resonant state. Locating its precise excitation energy and understanding the pronounced mirror asymmetry relative to its partner $^{17}$N, driven by the Thomas-Ehrman shift, have proven difficult~\citep{Garrido2004}. \\
In this work, we present a comprehensive theoretical study of the structure of $^{17}$Ne and its reaction on a heavy target. The structure of $^{17}$Ne is described within a three-body ($^{15}$O~+~p~+~p) framework using the pseudostates (PS) method, where the continuum is discretized using a transformed harmonic oscillator (THO) basis~\citep{PhysRevC.72.024007}. This basis suits weakly bound systems because it preserves the correct asymptotic behavior for bound states. \\
A major reaction mechanism for weakly bound nuclei is the breakup channel, 
in which the projectile dissociates into its constituents. Two types of breakup processes can be distinguished: elastic breakup (EBU), in which the target remains in its ground state and all projectile fragments emerge in the final channel, and non-elastic breakup (NEB), which encompasses processes in which one or more projectile fragments interact non-elastically with the target (including processes such as transfer of particles, incomplete fusion, and target excitation). The combined contributions of EBU and NEB are required to reproduce the experimental data in an inclusive breakup experiment.

To calculate the breakup and elastic scattering cross sections, we apply the continuum-discretized coupled-channel (CDCC)~\citep{PhysRevC.77.064609} formalism for the $^{17}$Ne~+~$^{208}$Pb reaction at an energy around the Coulomb barrier. Our theoretical results are directly compared with experimental data for elastic and breakup cross sections measured at the GANIL facility at the E$_{\rm{lab}}=136$ MeV~\cite{OVEJAS2023138007}. This comparison allows us to test the validity of the three-body structural model and to assess the effects of continuum couplings on the reaction mechanism.

The paper is structured as follows: Section 2 details the three-body model for the \texorpdfstring{$^{17}$Ne}{17Ne} projectile, including the THO pseudostates formalism, and presents and compares the CDCC results for the elastic and breakup channels with the GANIL data at \texorpdfstring{E$_{\rm{lab}}=136$ MeV}{Elab=136 MeV}. Section 3 discusses the physical implications of these results, including the role of the resonant states in reproducing the experimental data. Finally, a summary and conclusions are given in Sect. 4.


\section{Formalism}
\subsection{Structure}

In describing a three-body system, Jacobi coordinates $\{\mathbf{x}_k,\mathbf{y}_k\}$ are commonly used. The vector $\mathbf{x}_k$ denotes the relative coordinate of one pair, whereas $\mathbf{y}_k$ connects the center of mass of this pair to the third particle. On the basis of the Jacobi coordinates, one can introduce the hyperspherical coordinate set $\{\rho,\alpha_k,\hat{x}_k,\hat{y}_k\}$, where $\rho$ is the hyperradius, which is independent of the particular choice of Jacobi set, $\alpha_k$ is the hyperangle, and $\hat{x}_k$ and $\hat{y}_k$ are the angular coordinates associated with $\mathbf{x}_k$ and $\mathbf{y}_k$, respectively.
In the hyperspherical harmonics formalism, the total wavefunction of the three-body system can be written as
\begin{equation}
    \phi_{n}^{j\mu}(\rho,\Omega) = \dfrac{1}{\rho^{5/2}}\sum_{\beta} U_{n\beta}^{j\mu}(\rho)~\mathcal{Y}_{j\mu\beta}(\Omega)~\text{,}
    \label{eq1}
\end{equation}
where $\Omega\equiv\left\{\alpha_{k},\hat{x}_{k},\hat{y}_{k}\right\}$ and $\beta$ denotes the set of quantum numbers required to specify each channel, including the hypermomentum $K$. These quantum numbers arise from the couplings between the spin and orbital angular momenta of the three particles in the system. See Ref.~\cite{manoli} for details.\\
In Eq.~\ref{eq1}, $\mathcal{Y}_{j\mu\beta}(\Omega)$ are states of good total angular momentum, expanded in hyperspherical harmonics, whereas $U_{n\beta}^{j\mu}(\rho)$ are the hyperradial wave functions, obtained using the PS method~\cite{PhysRevC.72.024007}. Then, within this approximation, the hyperradial wavefunctions can be written as
\begin{equation}
  U_{n\beta}^{j\mu}(\rho)= \sum_{i}C_{n}^{ij\beta}~U^{THO}_{i\beta}(\rho)~\text{,}
    \label{eq2}
\end{equation}
where the maximum value of $i$ is the size of the hyper radial basis of the $\beta$ channel.
The PS method diagonalizes the three-body Hamiltonian in a discrete, square-integrable basis, where negative-energy eigenstates correspond to bound states and positive-energy eigenstates to continuum states. The most commonly employed discrete basis is the harmonic oscillator (HO) basis. However, the asymptotic behavior of HO functions does not coincide with the expected behavior of wavefunctions in weakly bound systems for bound states. One can use the THO method to address this inconsistency by applying a local scale transformation (LST)~\cite{PhysRevC.88.014327,PhysRevC.71.064601}. 
\subsubsection{Application to $^{17}$Ne}
    In this paper, $^{17}$Ne is represented as an $^{15}$O core plus two weakly bound protons.  Within the THO framework, the $^{17}$Ne nucleus is described in the Jacobi-T coordinate system, in which the two valence protons are connected through the $\mathbf{x}$ coordinate. For the $\rm p + {}^{15}\rm O$ interaction, an $l$-dependent potential, consisting of central and spin–spin components, was adjusted to reproduce the known resonant states of the unbound nucleus $^{16}$F. We adopted the same potential as in Ref.~\cite{Casal2016}. However, to reproduce the $5/2^-$ resonance, it was necessary to use a shallow potential in the s-wave channel instead of employing the adiabatic projection method to remove the Pauli-forbidden states.

\begin{table}[!ht]
\centering
\begin{tabular}{c c c } 
 \toprule
 \hline
 $\rm{\bf{j}}^{\pi}$ & $v_{\rm{3B}}~(\rm{MeV})$ & $\rm{\rho_{3B}}~\rm(fm)$  \\ 
 \hline
 $1/2^{-}$& 	1.8 & 5.0 \\ \hline
 $3/2^{-}$ & 1.3 & 5.0 \\ 
 \hline
 $5/2^{-}$ & 2.25 & 5.0  \\ 
 \hline
 $1/2^{+}$ & -5.02 & 5.0 \\ 
 \hline
 \bottomrule
\end{tabular}
\caption{Parameters of the three-body potential used to describe the structure of $^{17}$Ne.}
\label{Table1}
\end{table}

For the proton–proton interaction, the GPT potential~\cite{GOGNY1970591}, which includes central, spin–orbit, tensor, and Coulomb terms, was adopted. In addition to the two-body forces, a three-body potential of Gaussian form was used,
\begin{equation}
    V_{3\rm{B}}=v_{3\rm{B}}\exp{(-\rho/\rho_{3\rm{B}})}^{2}\text{.}
    \label{eq33}
\end{equation}
 The range parameter ($\rho_{3B}$) was kept fixed, whereas the potential depth ($v_{3B}$) was taken to be $\mathbf{j}^{\pi}$-dependent, as shown in Table~\ref{Table1}. It is worth mentioning that our calculations' component probabilities are in agreement with those in Ref.~\cite{GARRIDO200485}, for the $\mathbf{j}^{\pi}~=~1/2^{-},~3/2^{-},~5/2^{-}$ states, as can be seen in Table~\ref{Table2}.  

\begin{table}[!h]
\centering
\begin{tabular}{ c  c  c c c c} 
    \toprule
 \hline
 $\rm{\bf{j}}^{\pi}$ &$l_{x}$ & $l_{y}$ & L  & \multicolumn{2}{c}{Contribuition of the components}  \\ 
 
 &  & & & (This Work)&(\cite{GARRIDO200485})\\ \hline
  	$1/2^{-}$& 0 & 0 & 0 &  88\%& 89\%\\ \hline
   $3/2^{-}$ & 0  & 2& 2& 41\% &47.5\% \\ 
   & 2  & 0& 2& 18\% &25.7\% \\ \hline
   $5/2^{-}$  & 0 & 2& 2&  54\%& 47.6\% \\
     & 2 & 0& 2&  27\% & 25\%\\ \hline
      $1/2^{+}$  & 0 & 1& 1&  84\%& --  \\ \hline
     \bottomrule
\end{tabular}
\caption{Contributions of the dominant components to the wave functions of the low-lying states in $^{17}$Ne. The quantities $l_{x}$ and $l_{y}$ denote the orbital angular momenta associated with the Jacobi coordinates $x$ and $y$, respectively. The total orbital angular momentum is given by $\bf{L} = \bf{l}_x + \bf{l}_y$. The last two columns show the percentage contributions obtained in this work and those reported in Ref.~\cite{GARRIDO200485}, respectively.}
\label{Table2}
\end{table}
To check whether the structure model is suitable for describing $^ {17}$Ne, one can compute structure observables such as multipole transition probabilities. The  electric multipoles transition probabilities between the 1/2$^{-}$ ground state and the 3/2$^{-}$, 5/2$^{-}$, and 1/2$^{+}$ states was calculated, for a maximum size of the basis $\rm{i}_{max}=15$ and a maximum hypermomentum $\rm{K}_{max}=8$, following the prescription presented in Ref.~\cite{osti_1181614}. Table~\ref{Table3} presents the theoretical results from the present study, as well as those available in the literature, for the sum of $B(E\lambda)$ values associated with electric dipole and quadrupole transitions.
\begin{table}[!h]
\begin{tabular}{l c c c} 
 \toprule
 \hline
   & $\rm{B_{theo}(E}\lambda)$ & $\rm{B_{theo}(E}\lambda)$ & $\rm{B_{exp}(E}\lambda)$ \\ 
     Transition       &  $(\rm e^{2}fm^{2\lambda})$   &   $(\rm e^{2}fm^{2\lambda})$   & $(\rm e^{2}fm^{2\lambda})$    \\
           &   (This Work)  &  (Earlier Works)    &    \\
 \hline
  $B(E2)$ & \multicolumn{1}{c}{93.821}& 105~\cite{chromik2002two}	& $66^{+18}_{-25}$~\cite{chromik2002two}\\ 
  $1/2^{-}\rightarrow3/2^{-}$  &  &	&  \\ 
 \hline
 $B(E2)$ &128.122 & 155~\cite{chromik2002two}	&124(18)~\cite{chromik2002two} \\ 
   $1/2^{-}\rightarrow5/2^{-}$ &  &	& 90(18)~\cite{MARGANIEC2016200} \\ 
 \hline
 $B(E1)$  & 0.381 &	$2.5~\times~10^{-6}$~\cite{GRIGORENKO2006254} & -\\ 

 $1/2^{-}\rightarrow1/2^{+}$  &  &	 & \\ 
 \hline
 \bottomrule
\end{tabular}
\caption{\ensuremath{B(E\lambda)} values associated with electric dipole and quadrupole transitions.}
\label{Table3}
\end{table}

\begin{figure}[!h]
    \centering
    \includegraphics[width=7.0cm]{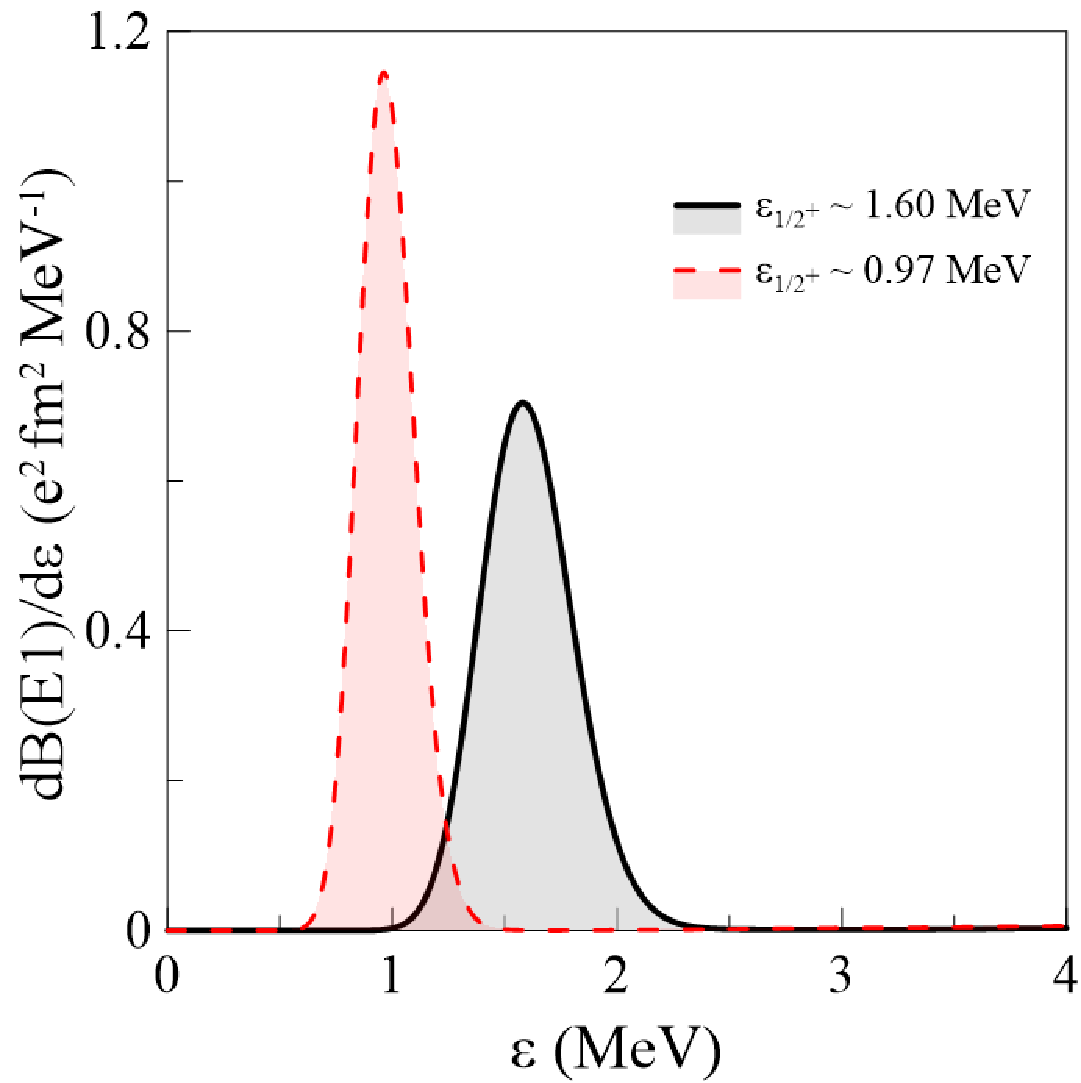}
    \caption{\label{fig:dbE1.eps}$B(E1)$ transition probability distribution from the 1/2$^{-}$ ground state to 1/2$^{+}$ continuum states in $^{17}$Ne using the THO method. }
\end{figure}

One observes a reasonable agreement between the results of the present work and those obtained in previous calculations, as well as the available experimental data, except for the 1/2$^{+}$ states, since there is no experimental data available for the electric dipole transition between the 1/2$^{-}$ and the 1/2$^{+}$ states. Furthermore, the value estimated in Ref.~\cite{GRIGORENKO2006254} is based on the $^{17}$N mirror nucleus, whereas in this work we calculate it using a three-body model for $^{17}$Ne.
The $B(E1)$ transition probability distribution from the 1/2$^{-}$ ground state to 1/2$^{+}$ states as a function of energy is presented in Fig.~\ref{fig:dbE1.eps}. We studied the effect of changes in the energy of the 1/2$^{+}$ resonant states on the $B(E1)$ distribution. In the first, the energy of the resonant state is fixed at the value suggested in the literature, $\rm{\varepsilon}_{1/2^{+}} \approx 0.97$~MeV above the breakup threshold, whereas in the second, the resonance energy is shifted to a higher value ($\rm{\varepsilon}_{1/2^{+}} \approx 1.60$~MeV). 

From this figure, it is clear that the shift in the resonance energy reduces the electric dipole transition distribution at low energies. This reduces the EBU cross section associated with these states in the four-body CDCC calculations, which is crucial for a reasonable description of the experimental EBU as well as the elastic cross sections around the Coulomb barrier, as will be discussed in the Sec.~\ref{sec:disc}.


\subsection{Four-body CDCC calculations}
From a theoretical point of view, a satisfactory treatment of reactions involving weakly bound nuclei must include a realistic description of the continuum. The CDCC approach is a reliable and well-established method for describing reactions involving these systems~\cite{AUSTERN1987125,Yahiro1986}. When the projectile is a three-body composite system, the method is an extension of the three-body CDCC formalism. The four-body CDCC formalism has been successfully applied to reactions involving Borromean weakly-bound projectiles, such as $^{6}$He, $^{9}$Be, $^{11}$Li, and so on~\citep{manoli,PhysRevC.73.051602,PhysRevC.92.054611,PhysRevC.97.044609,PhysRevLett.109.262701,PhysRevLett.110.142701}. 
The novel feature of the present work is the four-body CDCC calculation in the $^{17}$Ne~+~$^{208}$Pb reaction, in which the projectile is described as a three-body system.

\subsubsection{Application to \texorpdfstring{$^{17}$Ne~+~$^{208}$Pb at $E_{lab}=136$~MeV}{17Ne + 208Pb at Elab=136 MeV}}
\begin{figure}[!h]
    \centering
    \includegraphics[width=8.0cm]{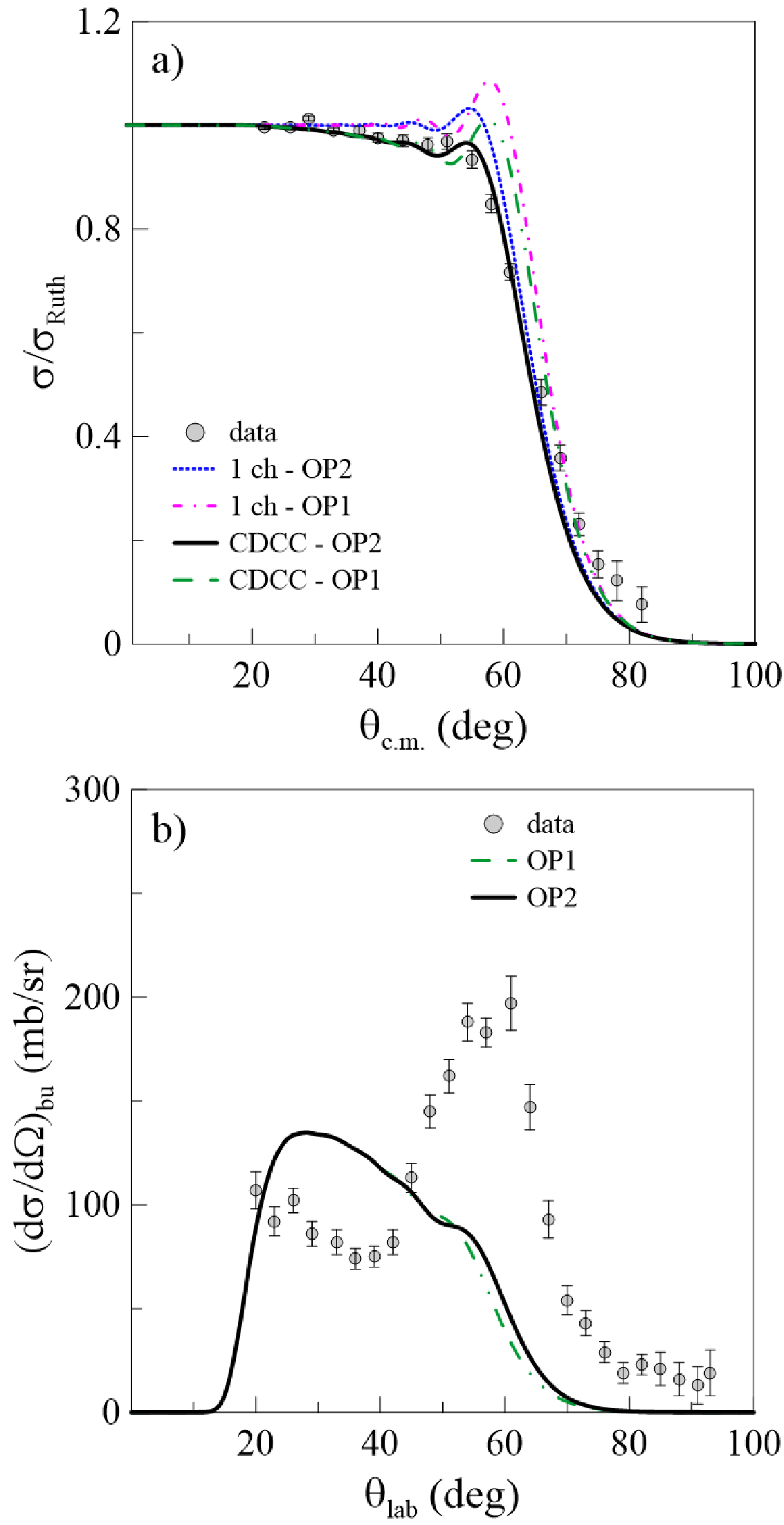}
    \caption{\label{fig:2elas-bu} Comparison of the one-channel and four-body CDCC results with the experimental data of Ref.~\cite{OVEJAS2023138007} for the elastic scattering (panel (a)), and for the EBU (panel (b)), showing the effect of doubling the depth of the imaginary part of the potential (OP1 $\rightarrow$ OP2).}
\end{figure}

In the present calculations, only the $^{17}$Ne states with $j^ {\pi}~=~1/2^{-},~3/2^{-},~5/2^{-}$, and $,1/2^{+}$ are considered. Within the CDCC formalism, these states, combined with $\rm{K}_{max}=8$ and $\rm{i}_{max}=15$, were sufficient to achieve converged elastic and breakup cross sections and to describe the experimental data for the $^{17}$Ne~+~$^{208}$Pb reaction of Ovejas \textit{et al.}~\cite{OVEJAS2023138007}.
For the p~+~$^{208}$Pb interaction, a Koning-Delaroche~\cite{KONING2003231} potential was used. For the $^{15}$O~+~$^{208}$Pb interaction, a phenomenological WS potential was chosen, including a central and Coulomb terms~\cite{BALL1975208}. This adopted potential was adjusted to reproduce the elastic scattering data of the $^{16}$O~+~$^{208}$Pb reaction. However, the $^{16}$O nucleus is doubly magic, so reactions involving it might have less absorption than expected for $^{15}$O. To check the effect of the imaginary part on the reaction observables, we used two different optical potentials. Optical potential 1 (OP1) is the one used in Ref.~\cite{BALL1975208}, and optical potential 2 (OP2) is identical to OP1, except that the depth of its imaginary part is twice that of OP1. Figures~\ref{fig:2elas-bu}~(a) and (b) illustrate the influence of the $^{15}$O~+~$^{208}$Pb interaction on the calculated observables. This point is examined further in Sec.~\ref{sec:disc}.


\begin{figure}[!h]
    \centering
    \includegraphics[width=8.0cm]{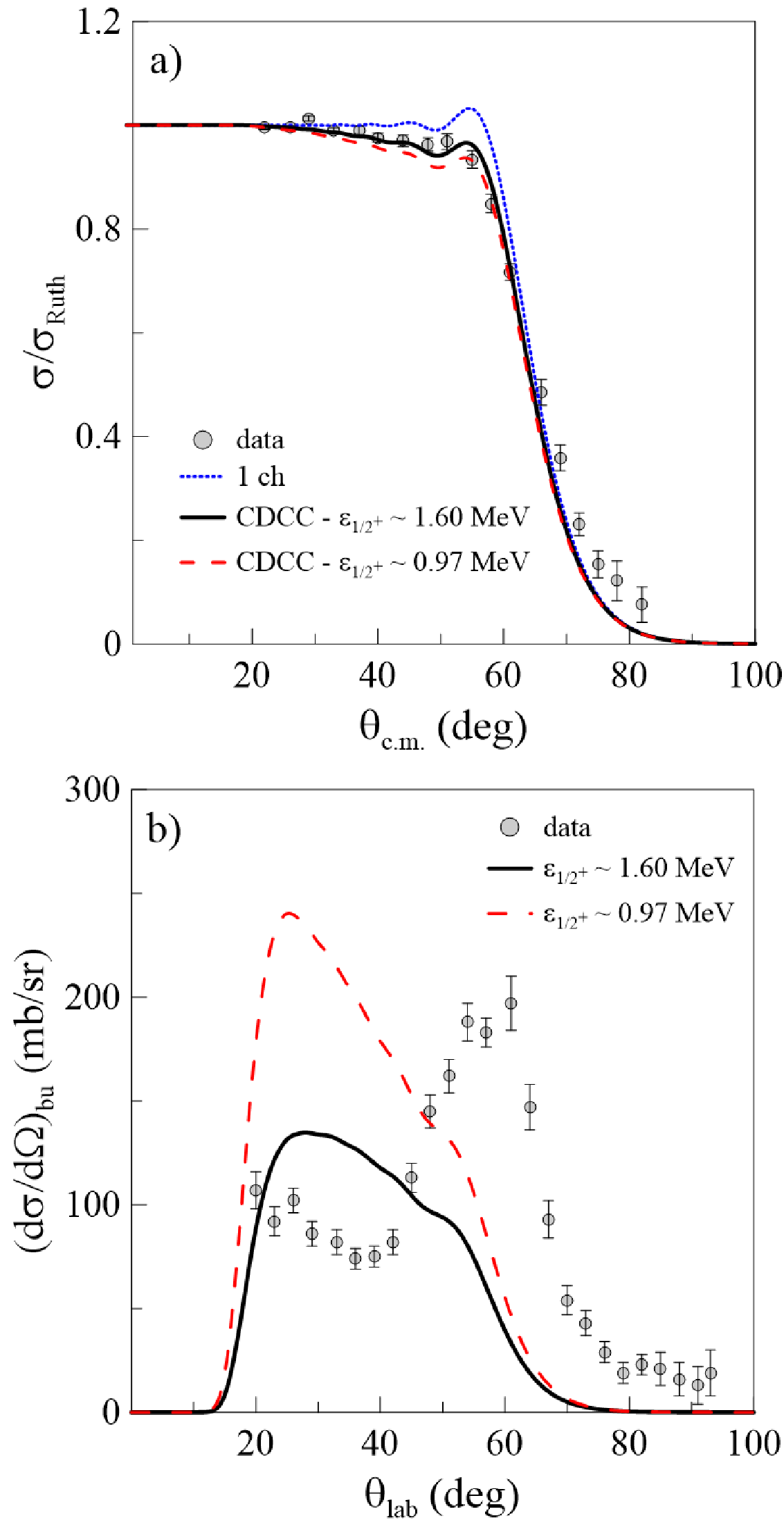}
    \caption{\label{fig:elasbu}Effect of shifting the 1/2$^+$ resonance energy from 0.97 to 1.6~MeV above the breakup threshold on the elastic scattering (panel (a)) and EBU (panel (b)) angular distributions.}
\end{figure}

Another interesting feature of this work is how changing the energy of the $1/2^{+}$ resonant state affects the observables. Figures~\ref{fig:elasbu}~(a) and (b) show the theoretical results for the elastic and breakup cross sections, respectively, in comparison with the corresponding experimental data, using the literature value of the $1/2^{+}$ resonance energy ($\varepsilon\approx 0.97$ MeV) and an adjusted value ($\varepsilon\approx 1.60$ MeV), both relative to the breakup threshold. The adjusted value provides better agreement with the experimental data. We note that the energy of this resonance is not well established in the literature because of its proximity to the $5/2^{-}$ state. This will be discussed in more detail in the next section.

\section{Discussion}
\label{sec:disc}
Figures~\ref{fig:2elas-bu}~(a) and (b) present the results obtained with both optical potentials (OP1 and OP2) used in the calculations. By comparing the one-channel calculation (i.e., neglecting the continuum) with the four-body CDCC result in Fig.~\ref{fig:2elas-bu}~(a), one can clearly observe the impact of the continuum states on the elastic cross section. As mentioned previously, the initially adopted optical potential (OP1) was constrained to reproduce the elastic scattering cross section for the $^{16}$O~+~$^{208}$Pb system. Since the $^{16}$O nucleus is doubly magic nucleus, reactions involving it might have less absorption than expected for $^{15}$O. In this way, our results indicate that using OP2 is essential to improve the agreement with the experimental elastic angular distribution for the $^{17}$Ne~+~$^{208}$Pb reaction and, in particular, to suppress the rainbow peak at $\theta = 60^{\circ}$.
Fig.~\ref{fig:2elas-bu}~(b) shows the corresponding impact of this modification on the calculated breakup cross section. The EBU cross section at forward angles remains unchanged, whereas the region around $\theta \approx 60^{\circ}$, associated with NEB processes, is slightly affected. To better understand the effect of changing the imaginary part of this potential, one can compute the integrated cross sections for the EBU and absorption mechanisms.

\begin{table}[!h]
\centering
\begin{tabular}{c c c} 
  \toprule
 \hline
 & $\sigma_{\rm{bu}}$ (mb) & $\sigma_{\rm{abs}}$ (mb) \\ 
 \hline
 OP1& 327.56  & 1477.6	  \\ \hline
 OP2&341.4& 1654.5 \\ \hline
\bottomrule
\end{tabular}
\caption{ Effect of doubling the imaginary part of the $^{15}$O~+~$^{208}$Pb potential on the EBU ($\sigma_{\rm{bu}}$) and absorption ($\sigma_{\rm{abs}}$) cross sections. The breakup cross section is practically unchanged, while the absorption cross section is significantly enhanced.}
\label{Table4}
\end{table}

As can be seen in Table~\ref{Table4}, the EBU cross section ($\sigma_{\rm{bu}}$) barely changes, while the absorption cross section ($\sigma_{\rm{abs}}$) increases by around 200~mb. This can be attributed to flux loss from the elastic channel to other channels not explicitly included in the coupling scheme.


Figures~\ref{fig:elasbu}~(a) and (b) show the effect of changing the energy of the 1/2$^+$ resonant states for the elastic and breakup cross sections. It is evident that this shift is essential for better reproducing the elastic scattering data at intermediate angles. This suggests that, within a three-body description of $^{17}$Ne, the 1/2$^+$ resonance plays a significant role in describing the experimental elastic cross section, alongside the 5/2$^{-}$ states. This is consistent with the findings of Ref.~\cite{Keeley:2023hia}, where the two-body model was unable to reproduce the $1/2^+$ resonance simultaneously with the $3/2^-$ and $5/2^-$ resonances. The best agreement with the experimental data was obtained by assuming a rather large $B(E2)$ value for the transition involving the $5/2^-$ and in particular, $3/2^-$ states. The authors suggested that a three-body description of $^{17}$Ne could account for the missing coupling strength.

This is in line with Ref.~\cite{Keeley:2023hia}, where a two-body model was unable to describe the $1/2^{+}$, $3/2^{-}$, and $5/2^{-}$ resonances consistently at the same time, and where the authors suggested that a three-body description of $^{17}$Ne could account for the missing coupling strength.

The shift also changes the EBU cross section, and including it is necessary for a reasonable description of the data as can be seen in Fig.~\ref{fig:elasbu}~(b). The peak at $\theta = 60^{\circ}$ is associated with NEB processes, which is beyond the scope of our model and therefore cannot be calculated here. A similar behavior was reported for $^{6}$He~\citep{Acosta:2011zz,ESCRIG20072}: at energies around the Coulomb barrier, the low binding energy of the projectile makes the backward-angle cross section dominated by processes other than EBU, such as transfer and incomplete fusion. This contrasts with $^{11}$Li, which is more weakly bound than both $^{17}$Ne and $^{6}$He, and whose angular distribution is dominated by EBU almost entirely~\citep{PhysRevLett.110.142701}.

Since the mirror nucleus $^{17}$N suggests that the dipole coupling should not be relevant for $^{17}$Ne~\cite{GRIGORENKO2006254}, one may ask whether the $5/2^-$ states alone could reproduce the elastic and EBU data. To test this, we lowered the energy of the $5/2^-$ resonance, making it narrower and strengthening its coupling to the ground state. Even in this case, the $5/2^-$ states by itself is unable to describe the data. Our calculations therefore indicate that the $1/2^+$ states is not only essential for reproducing the data, but that its position must also be shifted from 0.97 to 1.6~MeV above the breakup threshold.

\section{Summary and conclusions}
In this work, we have presented a theoretical study of the structure and reaction dynamics of the two-proton Borromean nucleus $^{17}$Ne on a heavy target. The bound and continuum states of $^{17}$Ne were described within a three-body ($^{15}$O~+~p~+~p) framework using the PS method in a THO basis. The three-body potentials were adjusted to reproduce the known experimental energy spectrum, including the $1/2^{-}$ ground state and the $3/2^{-}$, $5/2^{-}$, and $1/2^{+}$ resonant states, whose nature has been the subject of significant controversy. These states were then used as input for CDCC calculations of the $^{17}$Ne~+~$^{208}$Pb reaction at the near-barrier energy of 136 MeV, allowing direct comparison with experimental elastic and breakup cross sections measured at the GANIL facility.\\
Our analysis reveals several important findings. First, including continuum couplings significantly affects the elastic scattering cross section, showing that breakup channels cannot be neglected even when describing the elastic channel in this system. Second, to simultaneously describe the experimental data for both elastic and breakup cross sections, we found it necessary to shift the energy of the $1/2^{+}$ resonant state of $^{17}$Ne from 0.97 MeV to 1.6 MeV. This adjustment underlines the sensitivity of reaction observables to the spectroscopy of broad resonances near the drip line, where their precise energies and widths remain difficult to determine from structure calculations alone. Third, to suppress the rainbow peak that appeared at intermediate angles in the elastic cross section, the OP2 was used.  Importantly, this adjustment only affects the breakup angular distribution in the region associated with NEB processes, leaving the description of the EBU unchanged. This result emphasizes the critical role that the choice of optical potentials plays in the reaction dynamics of exotic nuclei with heavy targets.\\
In conclusion, this work presents the first four-body CDCC calculation for the  $^{17}$Ne~+~$^{208}$Pb system, treating the projectile consistently as a three-body ($^{15}$O~+~p~+~p) nucleus. This analysis demonstrates that a simultaneous description of elastic and breakup channels reveals a strong sensitivity to the position of the 
$1/2^{+}$ resonance in $^{17}$Ne, highlighting the interplay between structure and reaction dynamics in weakly-bound systems at Coulomb barrier energies. These findings motivate further experimental efforts to better constrain the low-lying continuum of $^{17}$Ne. Furthermore, this work provides a solid foundation for extending the four-body reaction approach to other astrophysically relevant reactions.
\section*{Acknowledgements}
Work supported in part by the Brazilian funding agencies, CNPq, FAPERJ, CAPES, and the INCT-FNA (Instituto Nacional de Ci\^encia e 
Tecnologia-F\'\i sica Nuclear e Aplica\c c\~oes), research project 408419/2024-5. 
This work is part
of the I+D+i Projects No. PID2023-146401NB-I00  and No. PID2024-159559NB-C22 funded by MCIN/AEI/10.13039/501100011033. 




\bibliographystyle{elsarticle-num} 
\bibliography{example}






\end{document}